\documentclass[12pt]{article}
\usepackage{graphicx}
\usepackage{rotating}
\usepackage{pdflscape}
\usepackage{color}
\usepackage{cite}

\def \beq{\begin{equation}}
\def \eeq{\end{equation}}
\def\eqref#1{(\ref{#1})}
\def\bea{\begin{eqnarray}}
\def\eea{\end{eqnarray}}

\def\URLtilde{\lower0.2em\hbox{$\tilde{\phantom{a}}$}}
\def\mycomm#1{\hfill\break\strut\kern-3em{\color{red}\tt ====> #1
\color{black}}\hfill\break}

\newcount\timecount
\newcount\hours \newcount\minutes  \newcount\temp \newcount\pmhours
\hours = \time
\divide\hours by 60
\temp = \hours
\multiply\temp by 60
\minutes = \time
\advance\minutes by -\temp
\def\hour{\the\hours}
\def\minute{\ifnum\minutes<10 0\the\minutes
\else\the\minutes\fi}
\def\clock{
\ifnum\hours=0 12:\minute\ AM
\else\ifnum\hours<12 \hour:\minute\ AM
\else\ifnum\hours=12 12:\minute\ PM
\else\ifnum\hours>12
\pmhours=\hours
\advance\pmhours by -12
\the\pmhours:\minute\ PM
\fi
\fi
\fi
\fi
}

\def\monthname{\relax\ifcase\month 0/\or January\or February\or
March\or April\or May\or June\or July\or August\or September\or
October\or November\or December\else\number\month/\fi}

\def\bold#1{\setbox0=\hbox{$#1$}     \kern-.025em\copy0\kern-\wd0
\kern.05em\copy0\kern-\wd0
\kern-.025em\raise.0433em\box0 }

\begin{document}
\setcounter{footnote}{1}
\rightline{EFI 17-16}
\rightline{TAUP 3022/17}
%\rightline{arXiv:1803.nnnnn}
\vskip1.5cm

\centerline{\large \bf Strange baryons with two heavy quarks}
\bigskip

\centerline{Marek Karliner$^a$\footnote{{\tt marek@proton.tau.ac.il}}
 and Jonathan L. Rosner$^b$\footnote{{\tt rosner@hep.uchicago.edu}}}
\medskip

\centerline{$^a$ {\it School of Physics and Astronomy}}
\centerline{\it Raymond and Beverly Sackler Faculty of Exact Sciences}
\centerline{\it Tel Aviv University, Tel Aviv 69978, Israel}
\medskip

\centerline{$^b$ {\it Enrico Fermi Institute and Department of Physics}}
\centerline{\it University of Chicago, 5620 S. Ellis Avenue, Chicago, IL
60637, USA}
\bigskip
\strut

\begin{center}
ABSTRACT
\end{center}
\begin{quote}
The LHCb Experiment at CERN has observed a doubly-charmed baryon $\Xi_{cc}^{++}
=ccu$ with a mass of $3621.40 \pm 0.78$ MeV, consistent with many predictions.
We use the same methods that led us to predict $M(\Xi_{cc},~J^P{=}1/2^+) = 3627
\pm 12$ MeV and $M(\Xi_{cc}^*,~J^P{=}3/2^+) = 3690 \pm 12$ MeV to predict
$M(\Omega_{cc}^+,J^P{=}1/2^+) = 3692 \pm 16$ MeV and $M(\Omega_{cc}^*,~J^P{=}
3/2^+) = 3756 \pm 16$ MeV.  Production and decay are discussed briefly, and
predictions for $M(\Omega_{bc})$ and $M(\Omega_{bb})$ are included.
\end{quote}
\smallskip

\leftline{PACS codes: 14.20.Lq, 14.20.Mr, 12.40.Yx}
\medskip

%\draft

% This is Section I
\section{INTRODUCTION \label{sec:intro}}

The LHCb Experiment at CERN has observed a doubly-charmed baryon $\Xi_{cc}^{++}
= ccu$ with a mass of $3621.40 \pm 0.78$ MeV \cite{Aaij:2017ueg}.  This value
is consistent with several predictions, including our value of $3627 \pm 12$
MeV \cite{Karliner:2014gca,note1}.  It is more than 100 MeV above a candidate
$\Xi_{cc}^+$ for an isospin partner claimed by the SELEX Collaboration
\cite{SELEX}, but not seen by others.  Here we use similar methods to those in
Ref.\ \cite{Karliner:2014gca} and earlier works \cite{Karliner:2008sv} to
predict the mass of the ground-state $ccs$ state with spin-parity $J^P=1/2^+$
$M(\Omega_{cc}^+) = 3692 \pm 16$ MeV and its hyperfine partner with $J^P =
3/2^+$, $M(\Omega_{cc}^*) = 3756 \pm 16$ MeV.  Binding effects lead the
difference between the strange and nonstrange doubly charmed baryon masses
to be less than half the constituent-quark mass difference between the strange
and nonstrange light quarks.  These results were obtained using constituent-%
quark masses appropriate for baryons.  Use of quark masses universal for
baryons and mesons leads to $M(\Omega_{cc})$ and
$M(\Omega_{cc}^*)$ about 40 MeV higher, with similar systematic variations
expected for $\Omega_{bc}$ and $\Omega_{bb}$, due mostly to uncertainty in
how strongly a strange quark binds to a heavy diquark.

In Section \ref{sec:ext} we list contributions to $M(\Omega_{cc}^+)$ that are
straightforward extrapolations of the calculation of $M(\Xi_{cc})$.  The pair
of charmed quarks is treated as a $(cc)$ diquark antisymmetric (a $3^*$) in
color and hence symmetric in spin ($S=1$).  The difference in binding between
a $(cc)$ diquark and a strange quark in comparison with binding between $(cc)$
and a nonstrange quark is discussed in Sec.\ \ref{sec:bin}.
Results using quark masses appropriate for both mesons and baryons are treated
in Sec.\ \ref{sec:uni}.  Production and decay are treated briefly in Sec.\
\ref{sec:pd}, predictions for $M(\Omega_{bc})$ and $M(\Omega_{bb})$ are
presented in Sec.\ \ref{sec:bcsbbs}, while results and comments on other work
are collected in Sec.\ \ref{sec:res}.

% This is Section II 
\section{EXTRAPOLATIONS FROM $\Xi_{cc}$ PREDICTION \label{sec:ext}}

We compare the contributions to the $\Xi_{cc}$ mass studied in Ref.\
\cite{Karliner:2014gca} to similar contributions to the $\Omega_{cc}$ mass in
Table \ref{tab:comp}.  We take quarks in a baryon to have effective masses
$m_q^b = 363$ MeV($q=u$ or $d$), $m_s^b = 538$ MeV, $m_c^b = 1710.5$
MeV, and $m_b^b=5043.5$ MeV.
We ignore isospin splitting, treated in \cite{Karliner:2017gml} and
references therein.
% This is Table I
\begin{table}
\caption{Comparison of contributions to the mass of the lightest doubly charmed
baryon $\Xi_{cc}$ \cite{Karliner:2014gca} with corresponding contributions to
the mass of $\Omega_{cc}$.
\label{tab:comp}}
\begin{center}
\begin{tabular}{c r c r} \hline \hline
\multicolumn{2}{c}{$\Xi_{cc} = ccq$} & \multicolumn{2}{c}{$\Omega_{cc} = ccs$}
\\
Contribution & Value (MeV) & Contribution & Value (MeV) \\ \hline
$2m^b_c + m^b_q$ & 3789.0 & $2m^b_c + m^b_s$ & 3959.0 \\
$cc$ binding & ${-}129.0$ & $cc$ binding & ${-}129.0$ \\
$a_{cc}/(m^b_c)^2$ & 14.2 & $a_{cc}/(m^b_c)^2$ & 14.2  \\
${-}4a/m^b_q m^b_c$ & ${-}42.4$ & ${-}4a'/m^b_s m^b_c$ & ${-}42.4$ \\
Total & 3626.8$\phantom{.0}\pm 12$ & Subtotal & 3801.8$\phantom{.0}\pm 12$ \\
\hline \hline
\end{tabular}
\end{center}
\end{table}

The effect of the spin-spin interaction between the $q$ quark and
the $(cc)$ diquark is parametrized by a term ${-}4a/m^b_q m^b_c$ , while
that between $s$ and $(cc)$ is parametrized by ${-}4a'/m^b_s m^b_c$
with $a' = a~m_s^b/m_q^b$ taken so that the two terms have the same strength.
This is motivated by comparing the spin-spin interaction in the $c \bar s$
and $c \bar q$ systems:  $M(D^*_s) - M(D_s) = 143.8$ MeV is almost the same
as $M(D^*) - M(D) =141.4$ MeV.  The smaller magnetic moment of $s$ is
compensated by a larger wave function at the origin in the $c \bar s$ system.
We assume a similar compensation is taking place here.  For the mass of the
$\Omega_{cc}^*(J^P = 3/2^+)$, we replace the term ${-}4a'/m^b_s m^b_c =
{-}42.4$ MeV by ${+}2a'/m^b_s m^b_c = {+}21.2$ MeV, so $M[\Omega_{cc}^*(J^P =
3/2^+)] - M[\Omega_{cc} (J^P = 1/2^+)] = 63.6$ MeV.

Our calculations of the masses of light hadrons, based on the ideas of Ref.\
\cite{DeRujula:1975ge}, use constituent-quark masses and do not require
separate binding energies.  However, for systems without $q$ involving heavy
quarks one must take into account additional binding.  For example, when
calculating the mass of the S-wave $c \bar s$ system, it was found necessary 
to include a supplemental binding energy of 69.9 MeV, while a binding energy of
258 MeV was needed to describe S-wave charmonium \cite{Karliner:2014gca}.
Hence the last energy in Table \ref{tab:comp} represents a subtotal; we
estimate the binding energy of $s$ with the diquark $(cc)$ in the next section.

% This is Section III
\section{DIQUARK--LIGHT QUARK BINDING \label{sec:bin}}

We shall interpolate between the $\bar c s$ and $\bar c c$ binding energies
to find that between $(cc)$ and $s$.  All three cases involve the
interaction of a color antitriplet with a color triplet.  We compare the
masses $m_1,m_2$ of the constituents and reduced mass $\mu \equiv m_1m_2/
(m_1+m_2)$ of the composite system in Table \ref{tab:m}.  When discussing
mesons, we use effective masses $m_q^m = 310$~MeV, $m_s^m = 483$ MeV, $m_c^m
= 1663.3$ MeV and $m_b^m=5003.8$ MeV\cite{Karliner:2014gca}.  
The mass of the $cc$ diquark is
calculated to be $2m^b_c - B(cc) + a_{cc}/(m^b_c)^2 = 3421.0 -129.0 + 14.2 =
3306.2$ MeV.  For use in subsequent discussion of the masses of $\Omega_{bc}
\equiv bcs$ and $\Omega_{bb} \equiv bss$, we include the binding energies
between the diquark $(bc)$ and $s$ and between the diquark $(bb)$ and $s$.
The mass of the $bc$ diquark is calculated to be $m_b^b + m_c^b - B(bc) =
5043.5 + 1710.5 - (167.6 \pm 3) = 6586.4 \pm 3$ MeV, where the error reflects
uncertainty in the $bc$ binding energy.  As the mass eigenstates are of
indefinite $bc$ spin (rather, they are approximately states of definite $cs$
spin), we ignore small hyperfine effects.  The mass of the $bb$ diquark is
calculated to be $2 m_b^b - B(bb) + a_{bb}/(m^b_b)^2 = 10087.0 - 281.4 + 7.8 =
9813.4$ MeV.
 
% This is Table II
\begin{table}
\caption{Comparison of constituent masses and reduced masses in MeV for some
systems of strange and $c$ or $b$ quarks, in a scheme with separate quark
masses for mesons and baryons. Binding energies in MeV are also shown, with
two different values averaged and errors reflecting half their difference for
the $(cc)s$, $(bc)s$, and $(bb)s$ systems (see text).
\label{tab:m}}
\begin{center}
\begin{tabular}{c c c c c} \hline \hline
  System   & $m_1$  & $m_2$  & $\mu$ &  $B$  \\ \hline
$\bar c s$ & 1663.3 &  483   & 374.3 &  69.9 \\
$\bar c c$ & 1663.3 & 1663.3 & 831.6 &  258  \\
  $(cc)s$  & 3306.2 &  538   & 462.7 & 109.4$\pm$10.5 \\
  $(bc)s$ & 6586.4$^a$ & 538 & 497.4 & 124.1$\pm$12.8 \\
  $(bb)s$ & 9813.4  &  538   & 510.0 & 129.4$\pm$13.4 \\ \hline \hline
\end{tabular}
\end{center}
\leftline{$^a$Mass eigenstates of indefinite $bc$ spin; small hyperfine terms
ignored}
\end{table}

The reduced mass of the $(cc)s$ system lies between those of $\bar c s$ and
$\bar c c$.  Assuming a power-law dependence on $\mu$, $B = A \mu^p$, gives
$p = 1.636$ and $B((cc)s) = 98.9$ MeV.  An alternate method makes use of the
Feynman-Hellmann
theorem \cite{FH}, which relates the derivative of an energy expectation value
with respect to a parameter $\mu$ to the expectation value of the derivative of
the Hamiltonian:
\beq
\frac{dE_\mu}{d\mu} = \langle \frac{dH_\mu}{d\mu} \rangle~.
\eeq
In the present case, the right-hand side is $-(1/\mu)\langle T \rangle$,
where $T$ is the kinetic energy.  Let us now assume $\langle T \rangle$ is
independent of the reduced mass.  This is indeed the case for a logarithmic
potential \cite{Quigg:1977dd,Quigg:1979vr}, which has been shown to suitably
interpolate between charmonium and bottomonium.  We shall assume $T$ is
constant also for our interpolation.  Then the shift in binding energy between
a system with reduced mass $\mu_1$ and one with $\mu_2$ is
\beq \label{eqn:FH}
\Delta B = \langle T \rangle \int_{\mu_1}^{\mu_2} \frac{d\mu}{\mu} =
 \langle T \rangle \ln \frac{\mu_2}{\mu_1}~.
\eeq
The binding energy increases with increased reduced mass, as expected.  One
can determine $\langle T \rangle = 235.6$ MeV by comparing $\bar c s$ and
$\bar c c$ binding energies, yielding
\beq
B((cc)s) = B(\bar c s) + \langle T \rangle \ln \frac{462.7}{374.3} = 69.9 +
50 = 119.9~{\rm MeV}~.
\eeq
The average of the two determinations is $109.4 \pm 10.5$ MeV, where we take
the error to be half of their difference.  Similar methods apply to the
estimates of $B((bc)s)$ and $B((bb)s)$ quoted in Table \ref{tab:m}, where the
averages are those of the power-law (lesser value) and Feynman--Hellmann
(greater value) methods of interpolation.  Subtracting this from the subtotal
in Table \ref{tab:comp}, whose error was assumed to be the same as in the
calculations of $M(\Xi_{cc})$, and adding the error of $\pm 10.5$ MeV in
quadrature, we find $M(\Omega_{cc}) = 3692 \pm 16$ MeV, $M(\Omega_{cc}^*)
= 3756 \pm 16$ MeV.

% This is Section IV
\strut\vskip-1.3cm\strut
\section{UNIVERSAL QUARK MASSES \label{sec:uni}}

For many years it has been realized that fits to baryon masses require
constituent quarks about 55 MeV heavier than those in fits to low-lying mesons
\cite{Lipkin:1978eh,Gasiorowicz:1981jz}.  An alternative, secondary, description
\cite{Karliner:2016zzc} makes use of quark masses appropriate for both mesons
and baryons, adding a term $S = 165.1$ MeV to characterize the extra mass in a
baryon due to a string junction \cite{Rossi:2016szw}.  The contributions to
$M(\Omega_{cc})$, before accounting of binding between the $(cc)$ diquark and
the strange quark, are shown in Table \ref{tab:alt}.
Also shown are contributions to $M(\Xi_{cc})$ in this scheme.  Here $m_q =
308.5$ MeV, $m_s = 482.2$ MeV, $m_c = 1655.6$ MeV, and $m_b=4988.6$ MeV
\cite{Karliner:2016zzc}.

The $(cc)$ diquark's mass is $M(cc,3^*) = 2(1655,6) - 121.3 + 14.2 = 3204.1$
MeV.  To account for binding between the $s$ quark and the $(cc)$ diquark, we
interpolate as before, with the results shown in Table \ref{tab:mu}.  The
binding energy in the $\bar c s$ system has been calculated as $B(\bar c s)=
-[3M(D_s^*) + M(D_s)]/4 + m_s + m_c =-[3(2112.1) + 1968.3]/4 + 482.2 + 1655.6$
 MeV = 61.65 MeV. 

% MK Tables III and IV moved to follow the above paragraph
% This is Table III
\strut\vskip-1.0cm
\begin{table}[h]
\caption{Contributions to $M(\Omega_{cc})$ and $M(\Xi_{cc})$ in a picture with
identical quark masses for mesons and baryons.
$a_{uqm}$ and $a'_{uqm}$ denote the strengths of $cq$ and $cs$ color hyperfine
coupling appropriate for universal quark masses~\cite{Karliner:2016zzc}.
\label{tab:alt}}
\begin{center}
\begin{tabular}{c r c r} \hline \hline
\multicolumn{2}{c}{$M(\Omega_{cc})$} & \multicolumn{2}{c}{$M(\Xi_{cc})$} \\
Contribution & Value (MeV) & Contribution & Value (MeV) \\ \hline
$2m_c + m_s$ & 3793.4 & $2m_c + m_q$ & 3619.7 \\
$cc$ binding & ${-}121.3$ & $cc$ binding & ${-}121.3$ \\
     $S$     &  165.1 &      $S$     &  165.1 \\
$a_{cc}/(m_c)^2$ & 14.2 & $a_{cc}/(m_c)^2$ & 14.2 \\
${-}4a'_{uqm}/m_s m_c$ & ${-}37.6$ & ${-}4a_{uqm}/m_q m_c$ & ${-}37.6$ \\
Subtotal & 3813.8$\phantom{.0}\pm 12$ & Total & 3640.1$\phantom{.0}\pm 12$ \\
\hline \hline
\end{tabular}
\end{center}
\end{table}
\strut\vskip-2.2cm\strut
% This is Table IV
\begin{table}[h]
\caption{Constituent and reduced masses in MeV for interpolation to find
binding energy between $s$ and heavy diquarks $(cc)$, $(bc)$, and $(bb)$, in a
scheme with common quark masses for mesons and baryons.  For the heavy diquark
systems two different values have been averaged; errors reflect half their
difference.
\label{tab:mu}}
\begin{center}
\begin{tabular}{c c c c c} \hline \hline
  System   & $m_1$  & $m_2$  & $\mu$ &  $B$  \\ \hline
$\bar c s$ & 1655.6 &  482.2 & 373.4 &  61.65 \\
$\bar c c$ & 1655.6 & 1655.6 & 827.8 &  242.7$^a$ \\
  $(cc)s$  & 3204.1 &  482.2 & 419.1 & 81.6$\pm$6.4 \\
  $(bc)s$ & 6484.9$^b$ & 482.2 & 448.8 & 94.0$\pm$9.4 \\
  $(bb)s$ & 9718.9     & 482.2 & 459.4 & 98.4$\pm$10.4 \\ 
\hline \hline
\end{tabular}
\end{center}
\strut\vskip-0.9cm
 \leftline{\strut\qquad \qquad $^a$From Ref.~\cite{Karliner:2016zzc}}
 \leftline{\strut\qquad \qquad $^b$Mass eigenstates of indefinite $bc$ spin;
 small hyperfine terms ignored}
\end{table}
\clearpage
Interpolating via a power law with $B = A \mu^p$ one finds $p = 1.721$, B((cc)s)
 =75.2 MeV, while interpolating via the Feynman-Hellmann theorem (\ref{eqn:FH})
 one finds $\langle T \rangle = 227.4$ MeV and $B((cc)s) = 87.9$ MeV.  Hence
$B((cc)s) = 81.6\pm6.4$ MeV, implying $M(ccs, 1/2^+) = 3732\pm14$ MeV.  This is
40 MeV above the value we obtained with separate quark masses for meson and
baryons.  The uncertainty reflects in part the uncertainty in estimating the
binding energy between a strange quark and the heavy diquark.  A precise
measurement of $M(\Omega_{cc})$ could help distinguish between the two pictures
compared here.  We also quote the predicted value of $M(\Xi_{cc}) = 3640 \pm
12$ MeV in the scheme with universal quark masses.  This is not as close to
the experimental value as that in Ref.\ \cite{Karliner:2014gca}, but still
acceptable.  For the $\Omega_{cc}^*(J^P = 3/2^+)$ we replace the term
${-}4a'/m_s m_c = {-}37.6$ MeV in Table \ref{tab:alt} by ${+}2a'/m_s m_c = 
{+}18.8$ MeV, so we predict $M(\Omega_{cc}^*) - M(\Omega_{cc}) = 56.4$ MeV,
or $M(\Omega_{cc}^*) = 3789 \pm 16$ MeV.

In addition to the $(cc)s$ binding energy, Table~\ref{tab:mu} contains also
the $(bc)s$ and $(bb)s$ binding energies, obtained in an analogous way.
The latter are used in Sec. VI to predict the masses  $\Omega_{bc}$ and
$\Omega_{bb}$.

% This is Section V
\section{PRODUCTION AND DECAY \label{sec:pd}}

We can estimate the rate for production of $\Omega_{cc} = ccs$ by reference to
that for $\Xi_{cc}^{++} = ccu$.  Imagine that some process gives rise to the
$(cc)$ diquark, which then fragments into $\Xi_{cc}$ by picking up a $u$
quark.  The corresponding process giving rise to $\Omega_{cc}$ then involves
$(cc)$ picking up a $s$ quark.  What is the ratio of these two processes?

There is information on $b$ quark fragmentation in hadronic collisions from
the CDF Collaboration \cite{Aaltonen:2008zd}, which measures $f_s \simeq 0.3
f_u$ in $\bar p p$ collisions at $\sqrt{s} = 1.96$ TeV.  One could expect
a similar ratio for $(cc)$ to pick up a $u$ or $s$ quark.  At 13 TeV, in
a sample of $pp$ collisions consisting of an integrated luminosity of
1.7 fb$^{-1}$, the LHCb experiment accumulated $313 \pm 33$ $\Xi_{cc}^{++}$
events \cite{Aaij:2017ueg}.  One might then expect the same sample to contain
about $(100 \pm 10) R$ $\Omega_{cc}^+$ identifiable events, where $R$ is the
ratio of $\Omega_{cc}$ to $\Xi_{cc}$ decays into identifiable branching
fractions.

The $\Xi_{cc}$ was seen in the final state $\Lambda_c K^- \pi^+ \pi^+$.  One
decay process depicted in Fig.\ 1 of Ref.\ \cite{Aaij:2017ueg} involves the
initial $u$ and one of the initial charmed quarks $c$ in the $\Xi_{cc} = ccu$
ending up in the $\Lambda_c = ucd$.  If the initial baryon is $\Omega_{cc} =
ccs$, an initial $s$ and one of the initial charmed quarks will end up instead
in a $\Xi_c^0 = scd$.  The detectability of the $\Omega_{cc}$ will then
depend on the relative efficiencies for reconstruction of $\Xi_c^0$ and
$\Lambda_c$.

Another potentially useful decay mode of $\Xi_{cc}^{++}$ is into $\pi^+
\Xi_c^+$.  Its visibility at LHCb will depend on relative efficiencies for
reconstruction of $\Xi_c^+$ and $\Lambda_c$.  The corresponding decay mode of
$\Omega_{cc}$ is into $\pi^+ \Omega_c$.  The LHCb experiment has detected not
only the $\Omega_c$ but several excited states of it \cite{Aaij:2017nav} in the
final state $\Xi_c^+ K^-$, providing a test of ability to reconstruct $\Xi_c^+$.

% This is Section VI
\section{MASSES OF $\Omega_{bc} = bcs$ AND $\Omega_{bb} = bbs$
\label{sec:bcsbbs}}

We have seen that much of the uncertainty in prediction of $M(\Omega_{cc})$
lies in uncertainty of the binding energy between the $(cc)$ diquark
and the strange quark.  The same is true when predicting the masses of
$\Omega_{bc}$ and $\Omega_{bb}$. 
Extrapolating our results for nonstrange states \cite{Karliner:2014gca}
to ones in which $q = u,d$ is replaced with $s$, we take account of (1)
the $s-q$ mass difference, (2) differences in $((QQ')q)$ and $((QQ')s)$
binding, and (3) small differences in hyperfine splittings, to obtain the
results in Table \ref{tab:str}.  The use of universal quark masses raises
the prediction of all $\Omega_{Q_1Q_2}$ masses by about 40 MeV.

% This is Table V
\begin{table}
\caption{Summary of predictions of $\Omega_{QQ}$ masses, in MeV.  ``Separate''
denotes separate quark masses for mesons and baryons; ``universal'' denotes
universal quark masses for mesons and baryons.
\label{tab:str}}
\begin{center}
\begin{tabular}{c c c} \\ \hline \hline
                & Separate & Universal \\
$M(\Omega_{cc})$  &  3692$\pm$16 &  3732$\pm$14 \\
$M(\Omega_{bc})$  &  6968$\pm$19 &  7013$\pm$16 \\
$M(\Omega'_{bc})$ &  6984$\pm$19 &  7025$\pm$16 \\
$M(\Omega_{bb})$  & 10208$\pm$18 & 10255$\pm$16 \\ \hline \hline
\end{tabular}
\vskip-1.3cm\strut
\end{center}
\end{table}

% This is Section VII
\section{RESULTS \label{sec:res}}

Using the same methods used to obtain an accurate prediction of the mass of
the recently discovered doubly-charmed baryons $\Xi_{cc}^{++}$, we predict the
mass of its strange partner: $M(\Omega_{cc}) = 3692 \pm 16$ MeV. The hyperfine
partner of this state, with $J^P$ = $3/2^+$, is predicted to have a mass
$M(\Omega_{cc}^*) = 3756 \pm 16$ MeV.  Predictions for the ground state masses
of the $bcs$ baryons $\Omega_{bc}$ and $\Omega'_{bc}$ and the $bbs$ baryon
$\Omega_{bb}$ are also presented.
The use of universal quark masses with an added ``string-junction''
contribution for baryons raises these predictions by about 40 MeV.

\begin{landscape}
\begin{figure}
\begin{center}
\includegraphics[width=24cm]{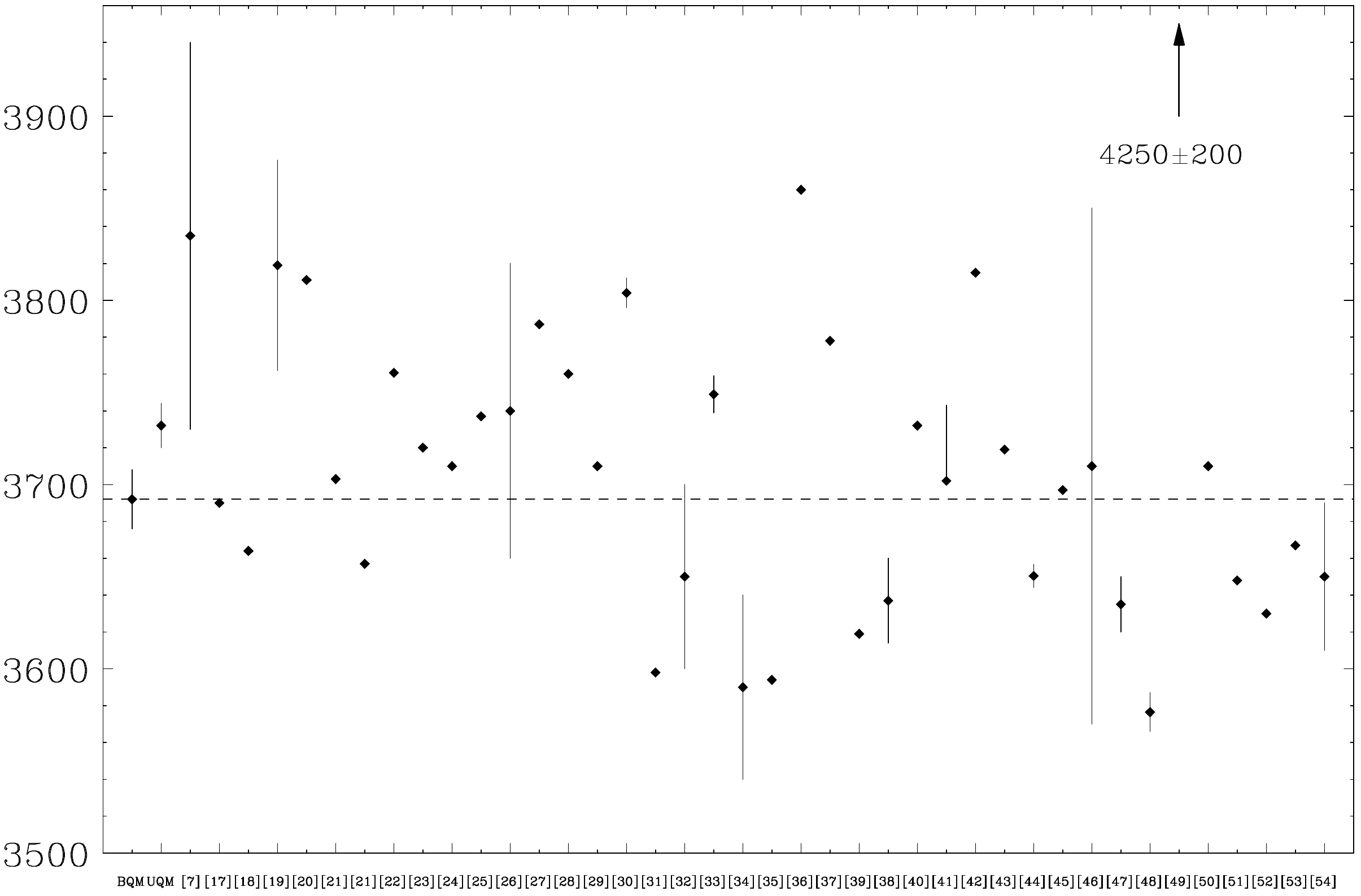}
\end{center}
\caption{Comparison of non-lattice predictions for $M(\Omega_{cc})$.  The
first two points are our predictions for baryonic quark mass (BQM;
dashed line) and universal quark masses (UQM).
\label{fig:omcc_comp1}}
\end{figure}
\end{landscape}

\begin{figure}
\begin{center}
\includegraphics[width=0.98\textwidth]{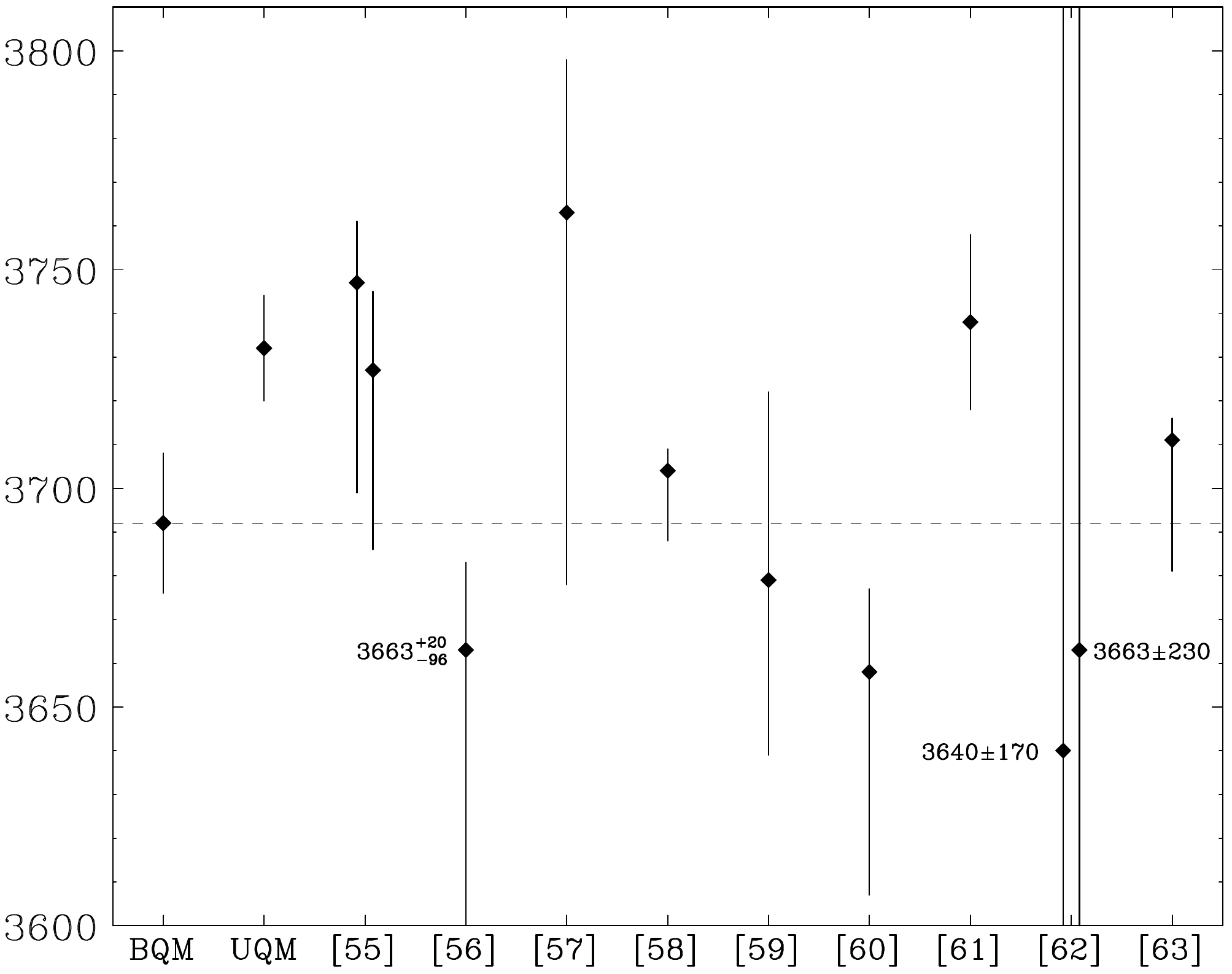}
\end{center}
\caption{Comparison of lattice predictions for $M(\Omega_{cc})$.  The first two
points are our (non-lattice) predictions for baryonic quark mass (BQM;
dashed line) and universal quark masses (UQM).
\label{fig:omcc_comp2}}
\end{figure}

Our predictions for $M(\Omega_{cc})$ are compared with a number of others in
Tables \ref{tab:omcc_comp1} (non-lattice) and \ref{tab:omcc_comp2} (lattice).
The predictions based on lattice gauge theory are shown separately as they have
less of a spread.  The corresponding values are plotted in Figs.\
\ref{fig:omcc_comp1} and~\ref{fig:omcc_comp2}.

In the picture with separate quark masses for mesons and baryons, the
prediction of a rather large value of $B((cc)s)$ distinguishes our
approach from a number of others \cite{note1} in which the difference
$M(\Omega_{cc}) - M(\Xi_{cc})$ is larger than our central value of 65 MeV.
In our calculation more than half of the the mass quark difference
$m_s^b - m_q^b = 175$ MeV is cancelled by increased binding.  For comparison,
a lattice gauge theory calculation \cite{Brown:2014ena} finds $M(\Xi_{cc}) =
3610(23)(22)$ MeV, $M(\Xi_{cc}^*) = 3692(28)(21)$ MeV, $M(\Omega_{cc}) =
3738(20)(20)$ MeV, $M(\Omega_{cc}^*) = 3820(20)(22)$ MeV, implying a difference
between the strange and nonstrange states of 128 MeV.  This is closer to the
value of 105 MeV we find in the picture with universal quark masses.

The production cross section for $\Omega_{cc}$ was estimated to be about
0.3 times that for $\Xi_{cc}^{++}$.  Its detectability then depends on the
relative efficiency for reconstructing $\Xi_c^0$ and $\Lambda_c$.

\strut\vskip-1.2cm\strut
\section*{ACKNOWLEDGMENTS}

We thank our LHCb colleagues for encouragement.
%The work of J.L.R. was supported by the U.S. Department of Energy, Division of
%High Energy Physics, Grant No.\ DE-FG02-13ER41958.
J.L.R. thanks Tel Aviv University for hospitality during completion of this
work.
% \clearpage
% This is Figure 1
% landscape environment to display the figure in landscape mode

% This is Figure 2
%
%force Fig 2 to be displayed before the Tables VI and VII
\clearpage

% This is Table VI
\begin{table}
\caption{Comparison of non-lattice predictions for $M(\Omega_{cc})$.  
\label{tab:omcc_comp1}}
\begin{center}
\begin{tabular}{c | c | c} \hline \hline
Reference & Value (MeV) & Method \\ \hline
Present work & $3692 \pm 16$ & Separate baryonic quark masses\\
Present work & $3732 \pm 14$ & Universal quark masses \\
\cite{DeRujula:1975ge}  & 3730--3940    & QCD-motivated quark model \\
\cite{Jaffe:1975us} & 3690 & Bag model \\
\cite{Ponce:1978gk} & 3664 & Bag model \\
\cite{Bjorken:1986xpa} & $3819 \pm 57$ & QCD-motivated quark model \\
\cite{Anikeev:2001rk} & 3811 & QCD-motivated quark model \\
\cite{Fleck:1989mb} & 3703 & Potential models\\
\cite{Fleck:1989mb} & 3657 & Bag models\\
\cite{Kerbikov:1987vx} & 3760.7$\pm$2.4$^a$ & Potential approach \\
\cite{Richard:1994ae} & 3720 & Potential model \\
\cite{Korner:1994nh} & 3710 & Heavy quark effective theory \\
\cite{Martin:1995vk} & 3737 & Potential model \\
\cite{Roncaglia:1995az} & $3740 \pm80$ & Feynman-Hellmann + semi-empirical \\
\cite{Lichtenberg:1995kg} & 3787 & Mass sum rules \\
\cite{Ebert:1996ec} & $3760 $ & Relativistic quasipotential quark model \\
\cite{SilvestreBrac:1996wp} & $3710 $ & Three-body Faddeev equations \\
\cite{Burakovsky:1997vm} & $3804\pm8$ & Quadratic mass relations \\
\cite{Gerasyuta:1999pc} & 3598 & Bootstrap quark model + Faddeev eqs.\\
\cite{Kiselev:2000jb} & $3650\pm50$ & Nonrelativistic QCD sum rules \\
\cite{Itoh:2000um} & $3749\pm 10$ &  Quark model\\
\cite{Kiselev:2001fw} & $3590 \pm 50$ & Potential approach + QCD sum rules\\
\cite{Kiselev:2002iy} & 3594 & Potential model \\
\cite{Narodetskii:2002ib} & $3860 $ & Nonperturbative string \\
\cite{Ebert:2002ig} & $3778 $ & Relativistic quark-diquark\\
\cite{He:2004px} & $3619 $ & Bag model \\
\cite{Chiu:2005zc} & $3637\pm23$ & Lattice; exact chiral symmetry \\
\cite{Migura:2006ep} & 3732 & Relativistic quark model + Bethe-Salpeter\\
\cite{Albertus:2006ya} & $3702^{+41}$ & Variational \\
\cite{Roberts:2007ni} & $3815 $ & Quark model \\
\cite{Martynenko:2007je} & 3719 & Relativistic quark model \\
\cite{Guo:2008he} & $3650.4 \pm 6.3^b$ & Quadratic mass relations \\
\cite{Valcarce:2008dr} & 3697 & Quark model + QCD \\
\cite{Wang:2010hs} & $3710\pm140$ & QCD sum rules \\
\cite{Weng:2010rb} & $ 3635 \pm 15 $ & Instantaneous approx.\ +
 Bethe-Salpeter\\
\cite{Patel:2008nv} & $3566 \div 3687$ & Potential model \\
\cite{Zhang:2008rt} & $4250 \pm 200$ & QCD sum rules \\
\cite{Bernotas:2008bu} & 3710 & Modified bag model \\
\cite{Giannuzzi:2009gh} & 3648 & Anti-de Sitter/QCD inspired potl.\\
\cite{Tang:2011fv} & 3630$^b$ & QCD sum rules \\
\cite{Ghalenovi:2014swa} & 3667 & Preferred potential model \\
\cite{Wei:2015gsa} & $3650\pm40^b$ & Quadratic mass relations \\
\hline \hline
\end{tabular}
\end{center}
\leftline{$^a$ Spin-weighted average of $M(\Omega_{cc})$ and $M(\Omega^*_{cc})$}
\leftline{$^b$ SELEX \cite{SELEX} $M(ccd,1/2^+) = 3519$ MeV candidate as input}
\end{table}

\clearpage

% This is Table VII
\begin{table}
\caption{Comparison of lattice predictions for $M(\Omega_{cc})$ with our
result.
\label{tab:omcc_comp2}}
\begin{center}
\begin{tabular}{c | c | c} \hline \hline
Reference & Value (MeV) & Method \\ \hline
Present work & $3692 \pm 16$ & Separate baryonic quark masses\\
Present work & $3732 \pm 14$ & Universal quark masses \\
\cite{Lewis:2001iz} &
\vrule width0pt depth1ex
$3747(9)({11\atop 47}) \div 3727(9)({16\atop 40})$ & Quenched lattice (LGT) \\
\cite{Flynn:2003vz} & 3663(11)(17)(95) & Quenched lattice\\
\cite{Liu:2009jc} & $3763 \pm 19 \pm 26^{+13}_{-79}$ 
  & Lattice, domain-wall + KS fermions\\
\cite{Namekawa:2013vu} & $3704(5)(16)$ & Lattice, $N_f=2+1$\\
\cite{Briceno:2012wt} & 3679(40)(17)(5) & LGT, $N_f=2+1$, $m_{\pi}=200$ MeV\\
\cite{Alexandrou:2014sha} & 3658(11)(16)(50)
 & LGT, $N_f=2+1$, $m_{\pi}=210$ MeV \\ 
\cite{Brown:2014ena} & 3738(20)(20) & Lattice \\
\cite{Sun:2016wzh} & $(3640\pm173)\div(3663\pm230)$ & Lattice; on-shell
renormalization \\
\cite{Alexandrou:2017xwd} & 3711(5)(30)& LGT, clover-improved, physical $m_\pi$
\\
\hline \hline
\end{tabular}
\end{center}
\end{table}

%
%\newpage

\end{document}